\documentclass[
 reprint,
 amsmath,amssymb,
 aps,
]{revtex4-2}
\usepackage{mathtools}
\usepackage{graphicx}
\usepackage{dcolumn}
\usepackage{bm}
\usepackage{siunitx}
\usepackage[utf8]{inputenc}
\usepackage[T1]{fontenc}
\newcommand{\mathbit}[1]{#1}
\makeatletter
\renewcommand*{\fnum@figure}{{\normalfont\bfseries \figurename~\thefigure}}
\renewcommand*{\@caption@fignum@sep}{\textbf{\usepackage{.} }}
\makeatother
\usepackage{amsmath}
\usepackage{setspace}
\usepackage {algpseudocode}
\usepackage{lipsum}

\makeatletter
\renewcommand*{\fnum@figure}{{\normalfont\bfseries \figurename~\thefigure}}
\renewcommand*{\@caption@fignum@sep}{\textbf{. }}
\makeatother
\renewcommand{\thefigure}{\arabic{figure}}

\begin{document}

\preprint{APS/123-QED}
\raggedbottom
\title{Deep Photonic Networks with Arbitrary and Broadband Functionality}

\author{Ali Najjar Amiri}
\email{aamiri20@ku.edu.tr}
\author{Aycan Deniz Vit}
\author{Kazim Gorgulu}
\author{Emir Salih Magden}
\email{Corresponding author: esmagden@ku.edu.tr}
\affiliation{%
 Department of Electrical and Electronics Engineering, Koç University\\
Sariyer, Istanbul, 34450, Turkey
}%


\begin{abstract}
Growing application space in optical communications, computing, and sensing continues to drive the need for high-performance integrated photonic components. Designing these on-chip systems with complex and application-specific functionality requires beyond what is possible with physical intuition, for which machine learning-based design methods have recently become popular. However, as the expensive computational requirements for physically accurate device simulations last a critical challenge, these methods typically remain limited in scalability and the optical design degrees of freedom they can provide for application-specific and arbitrary photonic integrated circuits. Here, we introduce a highly-scalable, physics-informed framework for the design of on-chip optical systems with arbitrary functionality based on a deep photonic network of custom-designed Mach-Zehnder interferometers. Using this framework, we design ultra-broadband power splitters and a spectral duplexer, each in less than two minutes, and demonstrate state-of-the-art experimental performance with less than 0.66 dB insertion loss and over 120 nm of 1-dB bandwidth for all devices. Our presented framework provides an essential tool with a tractable path towards the systematic design of large-scale photonic systems with custom and broadband power, phase, and dispersion profiles for use in multi-band optical applications including high-throughput communications, quantum information processing, and medical/biological sensing.
\end{abstract}

\maketitle


\section{Introduction}

Photonic integrated circuits (PICs) \cite{chrostowski2015silicon,bogaerts2018silicon} have significantly evolved over the last decade and are now essential technological components with critical importance in optical communications \cite{zhuang2015programmable}, sensing \cite{hu2017silicon,poulton2019long}, and computing \cite{zhang2020photonic,perez2017multipurpose,carolan2015universal}. With the growing diversity and complexity of photonic applications, designing custom PICs with state-of-the-art performance metrics has become one of the most critical drivers of advancement in photonic systems.

In traditional design of photonic components, the designer relies on prior knowledge of relevant system architectures, fundamental electromagnetic principles, and physical intuition to determine key parameters such as waveguide widths, lengths, or gaps  \cite{molesky2018inverse,piggott2015inverse,lu2013nanophotonic}. These parameters are then typically tuned either manually or by optimization algorithms, according to the results of repeated electromagnetic simulations, to achieve the required optical functionality. This approach yields a limited library of known devices and severely restricts the potential capabilities of the resulting photonic systems. More general approaches have emerged under the broad category of inverse/machine-optimized design \cite{piggott2015inverse,lu2013nanophotonic,qu2020inverse,tahersima2019deep,molesky2018inverse,zhang2022experimental}, allowing for greater design flexibility than manual tuning of waveguide parameters. Generally, these approaches perform more comprehensive searches over the complete domain of fabrication-compatible devices through heuristic \cite{zhang2013compact,sanchis2009highly} or gradient-based \cite{tahersima2019deep,jiang2021deep,so2020deep,hegde2020deep,jensen2011topology} computational optimization tools. Using these methods, devices such as couplers \cite{piggott2020inverse}, polarization splitters \cite{shen2015integrated,lu2013nanophotonic}, and spectral filters \cite{piggott2015inverse,zhang2022experimental} have been proposed and demonstrated. However, in these “free-form” design approaches, the degrees of design freedom are effectively controlled by the specified device footprint, which has key implications on the final device performance and the associated computational cost. While larger device footprints inherently provide the necessary design flexibility for complex and arbitrary optical functionality, they also rapidly scale the computational complexity of the iterative optimization process due to the physically-accurate electromagnetic simulations required   \cite{piggott2015inverse,lu2013nanophotonic,jia2018inverse,zhang2022experimental}. These requirements preclude the design of arbitrarily complex, ultra-broadband, or wavelength-specific photonic devices for the increasing number and variety of on-chip use cases and their custom application requirements.

The ideal approach to photonic design must allow for arbitrarily-specified photonic functionality while maintaining low computational cost. In recent years, programmable PICs made from Mach-Zehnder interferometers (MZIs) have been proposed as a potential solution to this problem  \cite{bogaerts2020programmable, xu2022self,perez2020multipurpose,zhuang2015programmable,capmany2020programmable}. These systems enable tuning of optical responses through thermal or electro-optic phase shifters in order to achieve wavelength-specific linear mappings between the input/output pairs  \cite{capmany2020programmable,shen2017deep,harris2018linear,miller2013self}. Several example applications including optical signal routing \cite{marpaung2019integrated,zhuang2015programmable,perez2020multipurpose}, image/signal classification \cite{shen2017deep,ashtiani2022chip,shastri2021photonics,feldmann2021parallel,lin2018all}, and quantum computing \cite{carolan2015universal,harris2016large} have already been demonstrated with numerous advantages such as post-fabrication reconfigurability, superior power efficiency, and lower latency over their electronic counterparts. However, the potential utility of photonic interferometer networks extends well beyond these demonstrated capabilities, with critical implications towards the design of photonic devices and systems with arbitrarily complex transfer functions.

In this paper, we introduce and experimentally demonstrate a highly-scalable framework for the design of photonic systems with arbitrarily-specified functionality, based on a deep photonic network architecture of custom-designed MZIs. Our architecture consists of a mesh of individually designed interferometers and is modeled by an equivalent computational network equipped with ultra-fast and physically-accurate simulation capabilities. In this network, each MZI is constructed from unique waveguide tapers, allowing for specific wavelength-dependent phase profiles to be achieved according to the target photonic functionality specified. The exact geometry of the individual interferometers is optimized by leveraging physics-informed machine learning capabilities in our design framework through a combination of rapid lookup of waveguide parameters and successive evaluation of photonic transfer matrices. Using this framework, we design ultra-broadband 50/50 and 75/25 power splitters and a spectral combiner/splitter, each in less than two minutes, with inherent fabrication compatibility on the 220-nm-thick silicon-on-insulator platform, and experimentally demonstrate state-of-the-art performance for all three devices. Our presented framework provides a path towards the systematic design of large-scale photonic systems with arbitrarily-specified, wavelength-dependent, or ultra-broadband responses.                      
\begin{figure*}
\includegraphics{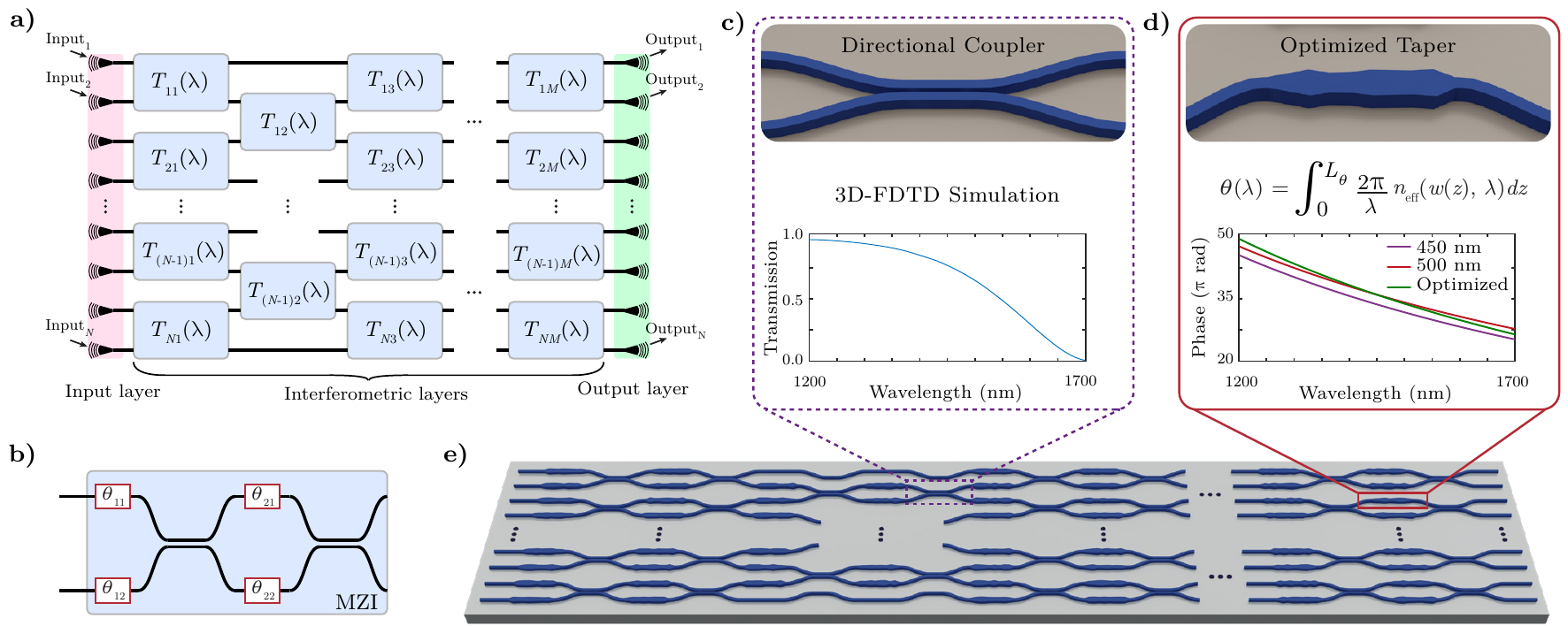}
\caption{\textbf{Deep photonic network architecture and components.} \textbf{(a)} The network architecture is composed of the input stage, horizontally-cascaded and vertically-repeated custom interferometric layers, and the output couplers. Each interferometric layer consists of a combination of Mach-Zehnder interferometers and individually-optimized waveguide structures. \textbf{(b)} Block diagram of a Mach-Zehnder interferometer with two pairs of waveguide tapers of custom geometries and two directional couplers. $\theta_{11}$ through $\theta_{22}$ indicate the phases  accumulated through each custom waveguide taper. \textbf{(c)} Schematic of the directional coupler with two S-bends and a 10 \si{\um}-long coupling section, and its 3D-FDTD simulated transmission response. \textbf{(d)} Schematic of an example custom waveguide taper constructed from a set of optimizable width parameters, from which the accumulated phase is calculated as a function of wavelength using the effective index. These custom waveguide tapers enable unique spectral phase profiles different from those in straight waveguides, as shown in the inset. \textbf{(e)} Overall structure of an example deep photonic network with cascaded interferometric layers of directional couplers and individually optimized waveguide tapers.}
\end{figure*}
\section{Results}
\subsection*{Deep Photonic Network Architecture}
The architecture of our deep photonic network consists of an input layer, a series of MZI layers, and an output layer, as shown in the schematic in Fig. 1(a). This architecture based on a mesh of MZIs has the theoretical capability to implement any linear $N \times N$ input-output mapping in order to achieve arbitrary optical functionality  \cite{reck1994experimental,miller2013self,miller2015perfect}. Input optical signal to the network is provided either externally by a series of couplers as shown, or by waveguides from upstream devices on-chip. The input optical signal is processed unidirectionally through layers of custom MZI interferometers, each with its own specific $2 \times 2$ mapping function denoted by $T_{ij}$. This modular network is modeled using the transfer matrix description of each one of its constituent building blocks, in a modular configuration. Specifically, each MZI consists of two pairs of waveguide tapers with custom geometries and two directional couplers, as illustrated in Fig. 1(b). The overall transfer matrix for each MZI is described by the transfer matrices of these constituent blocks as:
\begin{eqnarray}
T(\lambda) \tiny{=} 
\small{e^{-j\varphi(\lambda)}}
\begin{bmatrix}
t(\lambda) & -jq(\lambda) \\
-jq(\lambda) & t(\lambda) 
\end{bmatrix}
\begin{bmatrix}
e^{-j\theta_{21}(\lambda)} & 0 \\
0 & e^{-j\theta_{22}(\lambda)} 
\end{bmatrix}\nonumber\\
\times
\small{e^{-j\varphi(\lambda)}}
\begin{bmatrix}
t(\lambda) & -jq(\lambda) \\
-jq(\lambda) & t(\lambda) 
\end{bmatrix}
\begin{bmatrix}
e^{-j\theta_{11}(\lambda)} & 0 \\
0 & e^{-j\theta_{12}(\lambda)} 
\end{bmatrix}
\end{eqnarray}
where $t(\lambda)$, $q(\lambda)$, and $\varphi(\lambda)$ are the through- and cross-port amplitude coefficients and the phase response of the directional couplers, and $\theta_{11}(\lambda)$ through $\theta_{22}(\lambda)$ are the phases accumulated in corresponding waveguide tapers. The wavelength dependence of each one of these parameters plays a critical role in achieving arbitrary optical functionality in our networks. The directional couplers used throughout the network are identical and are designed to be approximately $50\%$ couplers at 1550 nm (see Supplementary Section 1 for details). A schematic of this directional coupler and its simulated through-port transmission are shown in Fig. 1(c). In contrast, all waveguide tapers are unique and custom-designed using a set of width and length parameters, as illustrated in Fig. 1(d), which are determined through an iterative optimization algorithm. The phase accumulated through each custom waveguide taper is calculated as a differentiable function of these custom widths ($w_{i}$), taper length ($L_{\theta}$), and input wavelength ($\lambda$), using the waveguide effective index $n_\mathrm{eff}(w,\lambda)$. This unique implementation allows the network to achieve wavelength-dependent phase profiles different from that of a straight waveguide, as demonstrated in the inset of Fig. 1(d), enabling much higher degrees of freedom while maintaining the same device footprint. Our design framework then constructs the overall photonic integrated circuit through an arbitrary number of interferometric layers as shown in Fig. 1(e).
\subsection*{Simulation and Optimization of the Optical Response through the Network}
Propagation of the complex optical amplitude through the network is carried out by a computational graph mimicking the physical network architecture. At each wavelength, the optical transformation carried out by the mesh of interferometers between $N$ input channels and $N$ output channels is represented by a computational graph. This architecture calculates the wavelength-dependent linear scattering matrix $S(\lambda)$ of the entire deep photonic network according to 
\begin{equation}
    S(\lambda) = \prod_{(i,j) \ \epsilon \ \Gamma} T_{i,j}(\lambda)
\end{equation}
where $\Gamma$ indicates the specific ordering of $2 \times 2$ transfer matrices in the network. This computation involves integrating the waveguide effective index using the custom widths and lengths for each waveguide taper, and extracting the directional coupler through-port, cross-port coefficients, and phase response from the 3D-FDTD results. In order for our custom photonic networks to be optimized for user-defined optical functionality, these operations are implemented through a differentiable programming construct, enabling both fast parameter lookups and automatic calculation of relevant derivatives \cite{jax2018github}. For calculating $\theta(\lambda)$, we numerically integrate the effective index throughout the length of the custom tapers using data obtained from Silicon Photonics Toolkit \cite{silicon-photonics-toolkit2022github}, an open-source software package providing access to several important propagation-related parameters in silicon waveguides as functions of wavelength and waveguide width. The directional coupler coefficients are similarly extracted from a differentiable interpolation of its 3D-FDTD simulation results. The result of this computation yields the complete network transfer function with a high degree of physical accuracy including the wavelength-dependent mappings for each input-output pair.   
\begin{figure*}
\includegraphics{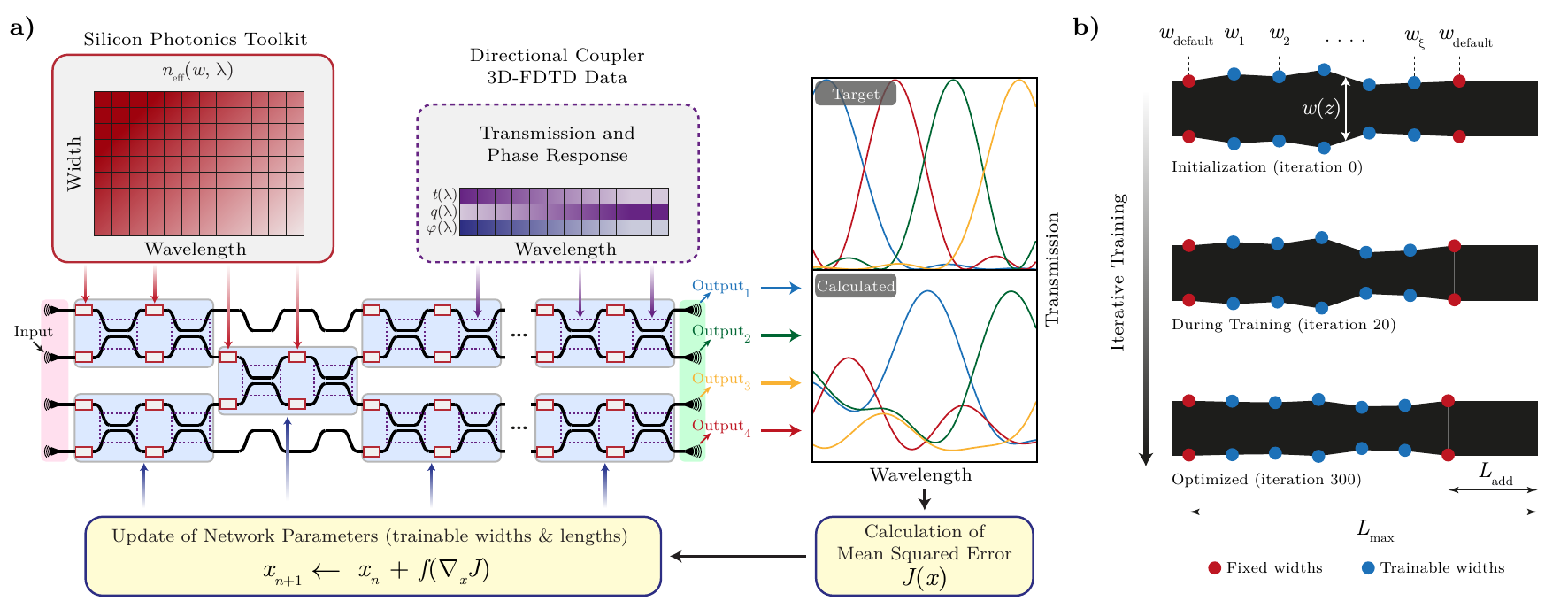}
\caption{\textbf{Optimization of an example 1-input 4-output photonic network.} \textbf{(a)} The $1 \times 4$ network structure is created with the desired number of layers, randomly-intialized custom waveguide tapers (red rectangles), and a target transmission response for input-output pairs. The mean squared error is computed from the difference between the calculated and target transfer functions, by summing over the specified wavelength range. The network parameters are trained iteratively through a backpropagation algorithm using the gradient of this error with respect to the design parameters denoted by $\mathbit{x}$ in the custom waveguide tapers. Other network components including the directional couplers and input/output layers are not trainable. \textbf{(b)} Evolution of a custom waveguide taper throughout optimization of the deep photonic network, where its geometry is shown at random initialization, at iteration 20, and at the end of optimization. Fixed widths ($w_\mathrm{default}$) and trainable widths ($w_1, w_2, …, w_{\xi}$) are marked with red and blue circles along the taper, respectively. At each iteration, an additional straight waveguide of length $L_\mathrm{add}$ is inserted at the end of the custom taper in order to achieve matching $L_\mathrm{max}$ lengths for all tapers.}
\end{figure*}
The ability to rapidly calculate a given network’s optical response as a differentiable function of its design parameters is critical from an optimization perspective. Using this capability, we construct an optimization procedure by iteratively modifying the waveguide tapers in order to obtain application-specific photonic networks with arbitrarily defined transfer functions. This procedure is illustrated for an example 1-input 4-output network in Fig. 2(a). First, we initialize a network with the desired number of interferometric layers and input-output ports. We define the target optical transfer function of these input-output pairs ($T_\mathrm{target}(\lambda)$), and assign semi-random width and length parameters to the constituent custom waveguide tapers. The network’s optical response is evaluated as a function of wavelength using the procedure described above and compared with the target transfer function. The difference between the calculated and target transfer functions is formulated as a mean squared error $J(\mathbit{x})=\frac{1}{Q}\sum_{\lambda}\lvert T_\mathrm{calculated}(\lambda , \mathbit{x})-T_\mathrm{target}(\lambda)\rvert^2$, where $Q$ is the number of wavelengths and $\mathbit{x}$ are design parameters including widths and lengths of the custom tapers. Gradient of $J(\mathbit{x})$ with respect to these design parameters $\nabla_{\mathbit{x}}J$ is calculated through a back-propagation procedure. We then minimize this error by iteratively modifying the widths and lengths of waveguide tapers, as illustrated in Fig. 2(b), using a gradient-based optimization algorithm \cite{kingma2014adam}. In addition to this error itself, we implemented numerous regularization schemes to achieve inherent fabrication compatibility by restricting waveguide widths from undergoing extreme changes in the custom tapers throughout the optimization procedure. Details regarding network initialization, convergence of this optimization process, and final resulting waveguide parameters can be found in Supplementary Section 2. 
\begin{figure*}
\includegraphics{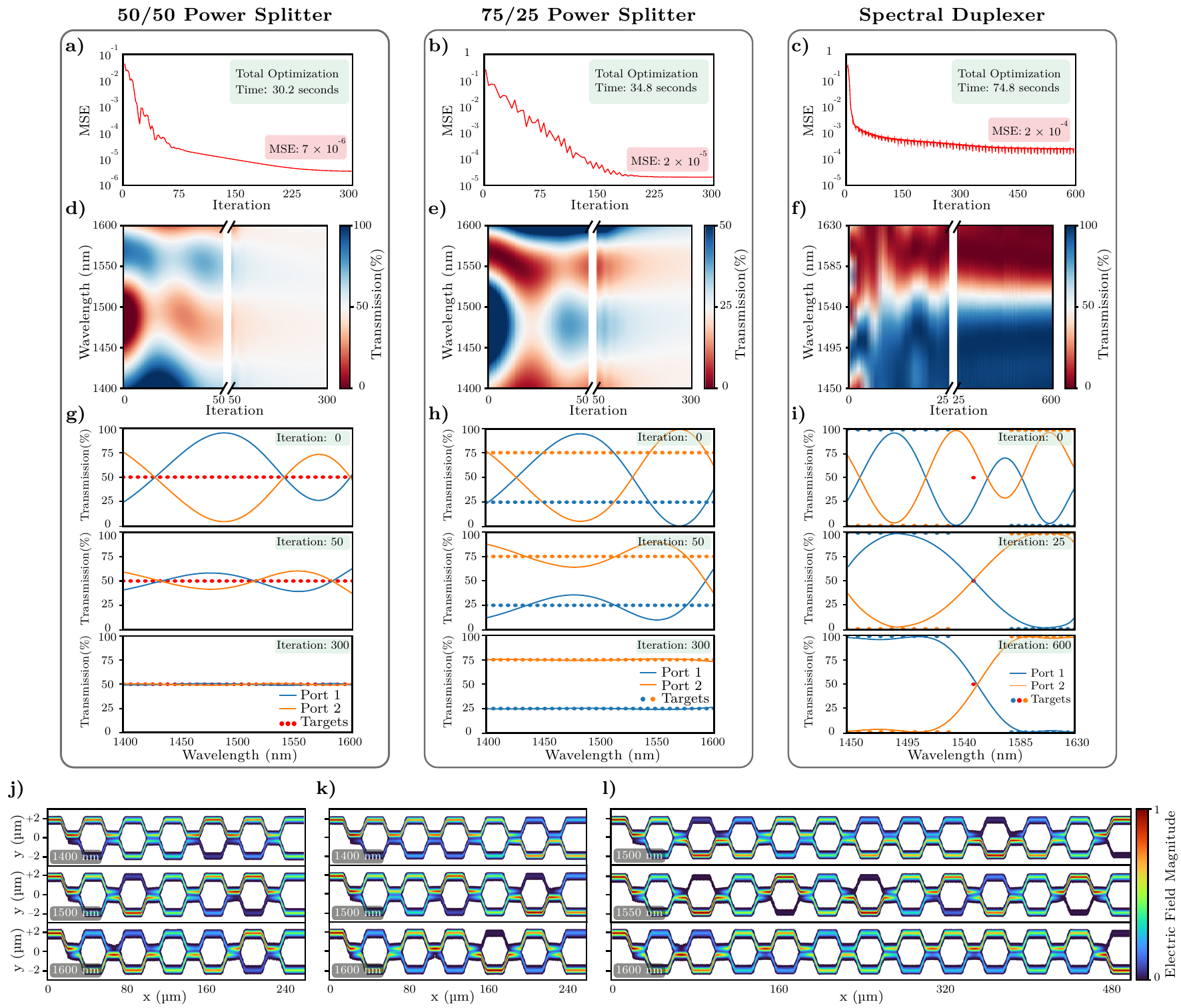}
\caption{\textbf{Optimization and final simulation results of power splitter and spectral duplexer deep photonic networks.} The mean squared error (MSE) versus iteration throughout optimization of \textbf{(a)} a 50/50 power splitter with 3 layers of MZIs (72 trainable parameters, 240 µm device length), \textbf{(b)} a 75/25 power splitter with 3 layers of MZIs (72 trainable parameters, 240 µm device length), and \textbf{(c)} a spectral duplexer with 6 layers of MZIs (144 trainable parameters, 480 µm device length). All three devices converge in several hundred iterations, within 1-2 minutes. \textbf{(d)-(f)} Transmission at the designated output port of each device as a function of wavelength. The evolution of this transmission through the iterative training process enables all three devices to achieve near-perfect transfer functions by the end of optimization. \textbf{(g)-(i)} Transmission spectra for each output during optimizations show gradual convergence to the target transfer functions indicated by the circles. The power splitters are optimized with 32 evenly-spaced wavelengths between 1400-1600 nm, and the duplexer is optimized with 21 wavelengths between 1450-1630 nm with a target cutoff at 1550 nm. Magnitude of the electric field at three different wavelengths obtained from 3D-FDTD simulations confirming broadband and flat-top operation for \textbf{(j)} the 50/50 power splitter, \textbf{(k)} the 75/25 power splitter, and \textbf{(l)} the spectral duplexer.}
\end{figure*}
\begin{figure*}
\includegraphics{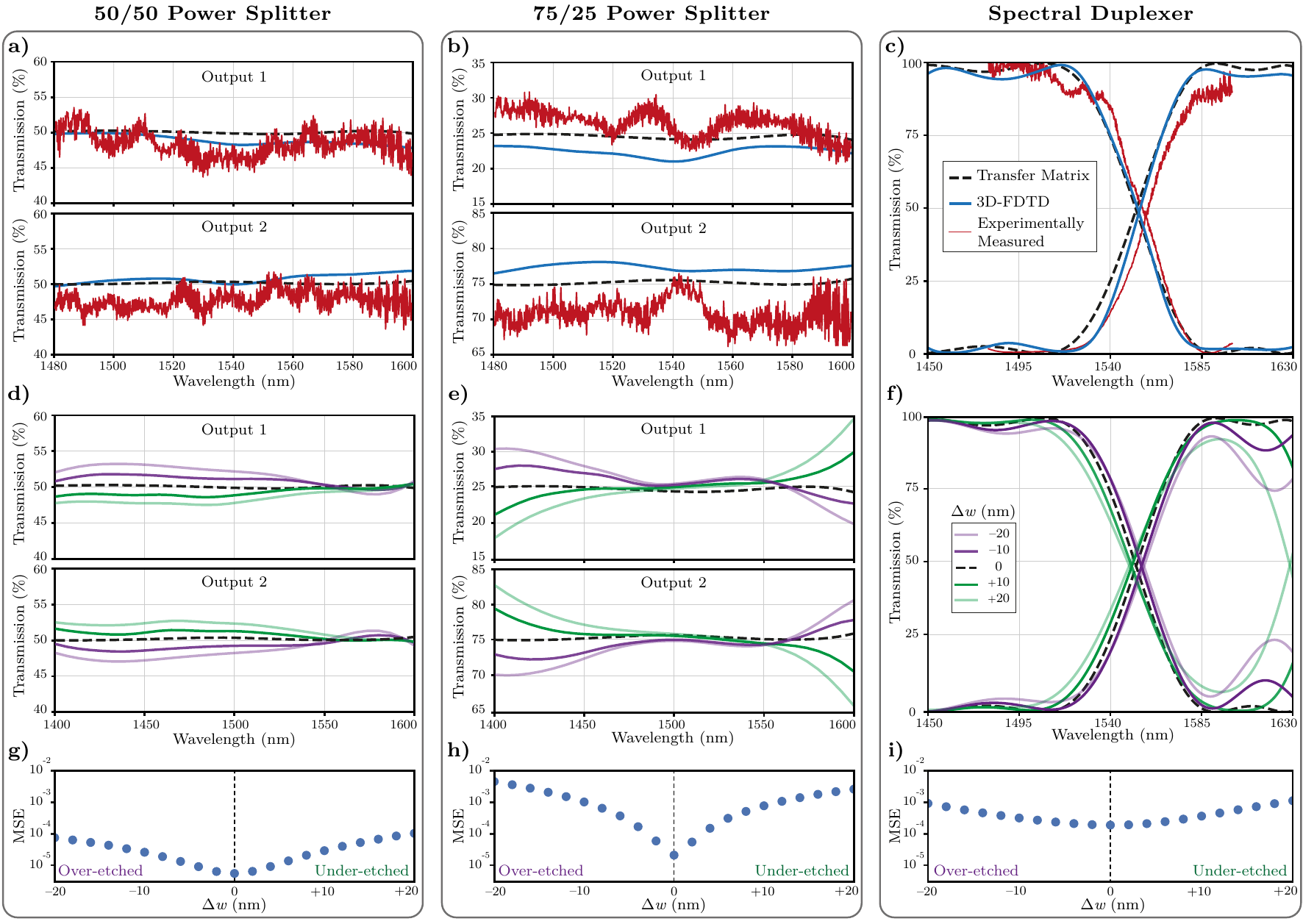}
\caption{\textbf{Experimental measurements and fabrication tolerance analysis of deep photonic networks.} \textbf{(a)-(c)} Measured transmission results together with transfer matrices and 3D-FDTD simulations at the output ports of the power splitters and the spectral duplexer. All three devices demonstrate agreement with simulation results over wide bandwidths with flat-top and low-loss transmission responses. \textbf{(d)-(f)} Transfer-matrix analysis of robustness against fabrication-induced variations for 10 nm and 20 nm over-etch and under-etch cases for the three devices. All components including directional couplers, S-bends, and waveguide tapers, are uniformly modified in simulation with the indicated etch offsets. \textbf{(g)-(i)} Resulting mean squared error in devices subject to over-etch and under-etch variations. With $±20$ nm modification of the waveguide widths, the resulting error typically increases by 1-2 orders of magnitude, corresponding to the changes in the simulated transfer function of the respective devices.}
\end{figure*}
\subsection*{Arbitrary Optical Functionality with Custom Deep Photonic Networks}
One of the key advantages of our proposed deep photonic network functionality is its ability to enable designs of photonic devices with arbitrary spectral specifications. We demonstrate how this capability allows for a universal design procedure for designing devices with ultra-broadband responses, and also devices with specific spectral features. As a proof of principle, this functionality is illustrated in Fig. 3 with three separate devices: two broadband power splitters with 50/50 and 75/25 splitting ratios operating within 1400-1600 nm, and a $1 \times 2$ spectral duplexer between 1450 nm and 1630 nm.  

Depending on the complexity of the desired functionality, our framework allows for the appropriate selection of hyperparameters of the deep photonic network including the number of interferometric layers and the number of custom widths in each waveguide taper. Details regarding the selection of hyperparameters can be found in Supplementary Section 3. Here, the power splitters are both designed with networks of three layers each, and the duplexer is designed with a network of six layers. For each custom waveguide taper in our devices, we used five trainable widths and a trainable length, resulting in a total of 24 parameters for each MZI in our photonic networks. The evolution of the resulting mean squared errors throughout the optimization processes are plotted in Fig. 3(a)-(c), where convergence is achieved in several hundred iterations and, at most, a few minutes on a single Tesla V100 GPU. Details regarding the optimization time of the photonic networks and their scalability can be found in Supplementary Section 4.

The wavelength-dependent design capability of our network is illustrated in Fig. 3(d)-(f), where we plot the transmission at one of the output ports for each one of the three devices as a function of wavelength. Throughout optimization, the output state evolves towards the target output functionality, as can be seen by the optical responses gradually approaching the desired 50\%, 25\% (for one output), and the spectrally duplexed outputs for the three devices, respectively. In Fig. 3(g)-(i), the transmission spectra at both output ports are plotted for each device at their randomly-initialized states at the beginning of optimization, at an intermediate state where the devices have been partially trained, and at the final states of the optimized devices. The final device responses demonstrate a near-perfect match with the specified target functionality. These responses are verified by the propagation of the optical input in the final optimized devices, which are plotted using the electric field intensity from 3D-FDTD simulations in Fig. 3(j)-(l). These simulation results confirm the expected outputs from our transfer matrix calculations that our networks are trained with. As expected, the power splitters achieve broadband operation; and the duplexer functions as a spectral splitter within its spectral design range, providing long-pass and short-pass outputs.   
\subsection*{Experimental Demonstration and Analysis of Deep Photonic Networks}
The fabricated power splitters and the spectral duplexer have been characterized using a continuous-wave tunable laser source (Santec TSL-710), an optical power meter (Santec MPM-210), on-chip grating couplers, and a polarization controller. The experimental characterization results for the two power splitters are shown in Fig. 4(a) and 4(b). For the 50/50 splitter, the maximum deviation from $50\%$ transmission is as low as $±6.42\%$ for both output ports; and the excess loss is measured to be less than 0.5 dB. As such, our network-based power splitter experimentally achieves a deviation of at most 0.6 dB within the 120 nm of measured bandwidth, and therefore a 1-dB bandwidth much wider than that. Similarly, for the 75/25 splitter, the deviations from the target transmission are within $±5.49\%$ ($±0.86$ dB) and $±8.88\%$ ($±0.55$ dB), for output ports number one and number two, respectively. The measured excess loss is less than 0.61 dB for both output ports. These results indicate that the 75/25 splitter achieves a 1-dB bandwidth of at least 120 nm, our widest measurement range possible. The spectral duplexer’s experimental characterization results are shown in Fig. 4(c). Within the pass-bands, a maximum loss of $11.45\%$ ($0.52$ dB) and $15.30\%$ ($0.72$ dB) are measured for the short-pass and long-pass outputs, respectively, and the excess insertion loss is measured about 0.66 dB (occurring at 1590 nm). The measured cutoff wavelength is around 1555.2 nm, compared to the specified target cutoff wavelength of 1550 nm. The extinction ratio between the two outputs is better than 15 dB for the majority of  the wavelength range characterized, and only reaches 13.6 dB at the edge of the measured spectrum (1600 nm). All three devices experimentally exhibit state-of-the-art performance and a close match with the training objective transmission responses. These results demonstrate and experimentally verify the universal capability of our design approach.   

Next, we analyze the robustness of our deep photonic networks against fabrication variations. Specifically, we plot the resulting transmission responses from transfer matrix calculations under potential over-etch and under-etch scenarios in Fig. 4(d)-(f) up to a change of $±20$ nm in the waveguide widths and gaps. The device responses are calculated by simulations of the network structures with updated waveguide tapers and directional couplers for the amounts of specified etch offsets. We observe minimal deviation of the transmission response from the ideal case with $±10$ nm over- and under-etch. At $±20$ nm, we observe more significant changes in the simulated transmission responses, resulting from changes in the wavelength-dependent phase profiles in waveguide tapers and the shifted responses of the directional couplers, as expected. This is also demonstrated in Fig. 4(g)-(i), where we plot the mean squared error of the resulting transmission with different over- and under-etch amounts. The calculated error increases with larger over-/under-etch amounts, indicating deteriorations in the resulting device performance. Functionally, we note that all three devices can still work as intended, with slightly inferior performance metrics up to the simulated $±20$ nm etch offsets.   
\subsection*{Deep Photonic Network Capability and Fabrication Robustness}
The scalability of our deep photonic networks and the computational efficiency of our underlying simulation/optimization framework can provide highly capable networks with extremely large degrees of freedom to design arbitrarily complicated optical devices. For this architecture, the selection of the number of interferometric layers is a major design choice that determines the number of degrees of freedom for the network. While the trainability and capability of the resulting network increase with the number of layers at first, each additional layer also introduces additional propagation loss due to the waveguide bends added with each layer. This tradeoff between device capability and excess loss can be modeled by analyzing devices with different numbers of layers trained for the same objective functionality. In Fig. 5(a), we plot the final mean squared error in the simulated transmission responses for different 50/50 power splitters designed with numbers of layers ranging from $M = 2$ to $M = 60$. As expected, the simulated error in the transmission response initially decreases and reaches a minimum with networks of 3 and 4 layers. However, with the increased number of layers, the accumulation of excess loss through additional layers outweighs the benefits of increased network capability, and results in a larger calculated error and an inferior transmission response.

Similarly, while longer networks with more interferometric layers can provide larger degrees of freedom and more complex optical capabilities, they are also less robust to fabrication variations. Similar to the added optical loss, errors in the phase profiles add up through the additional layers and negatively affect the resulting device performance. We analyze the fabrication tolerance of 50/50 splitters constructed from different numbers of layers in  Fig. 5(b), where the mean squared error is plotted as a function of the etch offset. The results demonstrate that longer networks (with greater numbers of layers) are more sensitive to fabrication-induced changes due to the accumulation of phase and coupling errors within consecutive MZIs. For instance, while the minimum error calculated is similar for 3-layer and 4-layer splitters, the 4-layer network exhibits significantly worse performance with etch offsets reaching $±$20 nm. This analysis serves as an important guideline towards determining the appropriate number of layers for the design of specific structures using the demonstrated custom networks. Similar analyses for the 75/25 power splitter and the spectral duplexer can be found in Supplementary Section 3.
\begin{figure}[hb]
\includegraphics{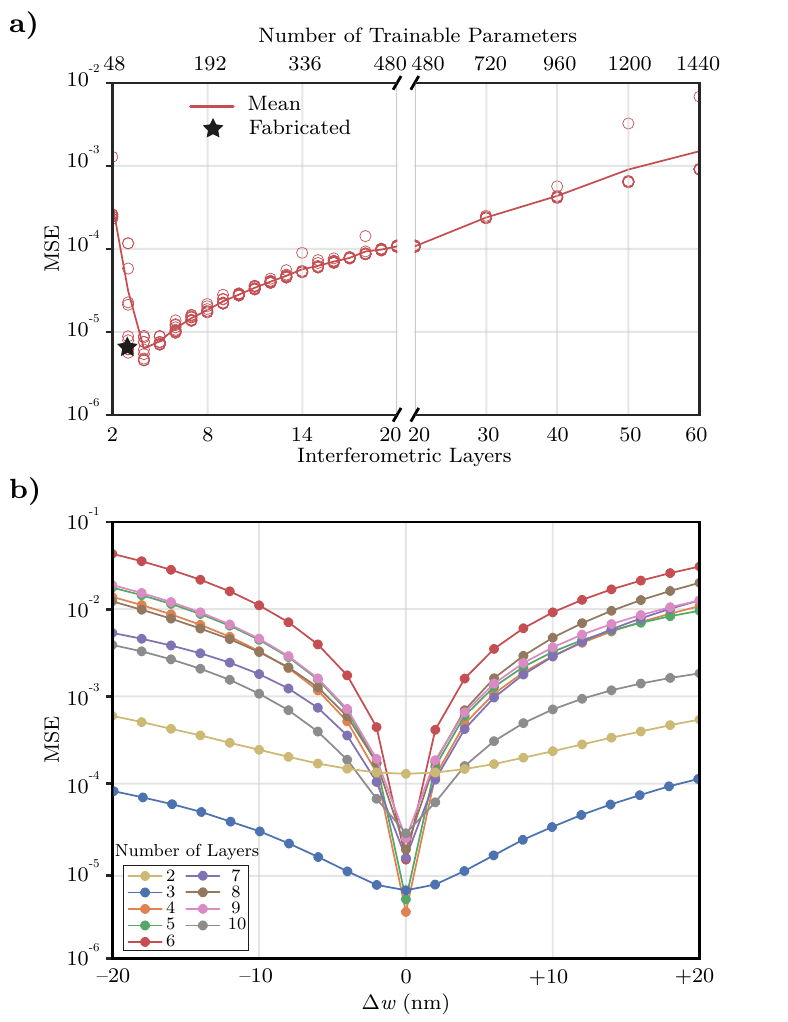}
\caption{\textbf{Influence of network size on final device performance.} \textbf{(a)} Performance of the $50/50$ power splitter deep photonic network as a function of the number of interferometric layers, which directly controls the number of trainable parameters. The plotted mean squared error includes propagation loss in the directional couplers and the S-bends extracted from their 3D-FDTD simulations. For each network size, ten different randomly-initialized devices are optimized and depicted with red circles. \textbf{(b)} Robustness of device performance against fabrication-induced variations with number of layers from $M=2$ through $M=10$. While increasing the number of interferometric layers initially provides better-performing devices under ideal fabrication conditions ($\Delta w=0$), longer devices perform worse under significant fabrication variations due to accumulating phase errors.}
\end{figure}
\section{Discussion}
Our design framework provides a computationally efficient, physically accurate, and systematic methodology for creating deep photonic network architectures for on-chip arbitrary optical systems. While we only demonstrated silicon-based devices, the presented methodology is applicable in a wide variety of material platforms and spectral applications. Despite having been designed by an arbitrarily capable photonic optimization framework, all three of our proof-of-concept devices experimentally demonstrate state-of-the-art performance. Even in comparison to structures with specific power splitter geometries \cite{kim2022experimental, Lin:19, Chang:18, chen2017broadband, Wang:16, Lu:15}, our network-type power splitters exhibit better experimental performance metrics with 1-dB bandwidths reaching as wide as the entire measured spectrum of 120 nm and over 200 nm in simulation. Similarly, our duplexer demonstrates better experimental performance than devices with similar functionality  \cite{zhang2022experimental, PhysRevA.100.023809, magden2018transmissive, piggott2015inverse}, with less than 0.66 dB insertion loss, flat-top transmissions at both outputs, and a cutoff wavelength shift of only 5 nm. Even though our networks at their current state do not implement reconfigurable photonic capabilities, they also do not require the use of electronic driving circuitry for detuning or calibration purposes. Consequently, our devices are not prone to long-term stability issues and excessive power draw requirements of systems with thermal, MEMS-based, or electro-optic phase shifters \cite{xu2022self, quack2023integrated}.

In summary, our design framework enables highly scalable implementations of arbitrary transfer functions on-chip, by casting the problem of photonic design as a constrained optimization problem with inherent fabrication compatibility. By integrating accurate waveguide parameters and 3D-FDTD simulations into a physics-informed machine learning architecture, this methodology enables rapid yet accurate simulations of photonic devices and their scalable optimization. Our modular network design allows for a large number of degrees of freedom through custom layers of MZIs, allowing for complex photonic functionality, and therefore presents a tractable path forward for the design of large-scale integrated photonic systems. Moreover, as our computational design framework keeps track of complete phase information through the individual network components, it allows for the design of photonic networks with specific phase and dispersion profiles as a part of their target functionality. Due to the availability of rapid individual device simulations, our framework can also be configured to enable future designs with on-chip amplifiers and lasers \cite{zhou2023prospects, li2018monolithically}, electrically-interfaced modulators and detectors \cite{liu2015review}, as well as structures with robustness against fabrication-induced variations \cite{perez2019scalable}. These capabilities present exciting novel directions in the design of photonic components with arbitrary transfer functions for use in next generation optical communication applications, neuromorphic photonic information processors, and medical/biological sensing. 
\section{Methods}
\subsection{Numerical Simulations}
The effective indices of silicon strip waveguides were extracted using Silicon Photonics Toolkit \cite{silicon-photonics-toolkit2022github}, an automatic differentiation-compatible open-source software package for the design of integrated photonic structures. This package enables fast lookup and evaluation of waveguide parameters on the 220 nm SOI platform, which is critically important for the rapid and scalable evaluation of our optical transfer functions. In our deep photonic networks, optical responses of the other components including directional couplers and waveguide bends were extracted from 3D-FDTD simulations performed with a maximum spatial discretization of 17 nm in all three dimensions. These responses including both amplitude and phase information were then linearly interpolated at 1000 wavelengths between 1.2 \si{\um} and 1.7 \si{\um}. The resulting interpolations were implemented as automatic differentiation-compatible lookup functions, and used during the performance evaluation of the constructed photonic networks. 
\subsection{Numerical Optimization Framework}
Our deep photonic network optimization framework was built on an open-source, end-to-end deep learning library \cite{Trax2020}, enabling the use of state-of-the-art machine learning software constructs as well as access to modern hardware accelerators including GPUs and TPUs. In this framework, we model each interferometric structure as part of a physics-informed artificial neural network, and evaluate the amplitude and phase profiles of the transfer functions between each input/output pair using the automatic differentiation-compatible functions described above. This modular and highly parallelizable architecture allows for serial, parallel, or even residual types of connections between interferometric layers, which can also be used for constructing more complicated network topologies. The trainable parameters of our networks are iteratively optimized using adaptive moment estimation \cite{kingma2014adam}. During the optimizations, the learning rate was progressively reduced from $3 \times 10^{-3}$ to $10^{-4}$ for ease and speed of convergence. For the design of the power splitters and the spectral duplexer, we used batch sizes of 32 and 21, respectively. A relative convergence was used for the stop condition of optimizations (see Supplementary Section 4 for details). All optimizations were performed using a single Tesla V100 GPU.   
\subsection{Device Fabrication}
After optimization, the final designed devices were converted to mask layouts using capabilities implemented in our design framework, through an open-source layout construction software library \cite{gdstk}. Grating couplers were added at the inputs and outputs of the network in order for on- and off-chip light coupling. The devices were fabricated using standard 193 nm CMOS photolithography techniques on the SOI platform with a 220-nm-thick silicon device layer through IMEC’s multi-project-wafer foundry service.

\begin{acknowledgments}
This work was supported by the Marie Sklodowska-Curie Fellowship (no 101032147) through the Horizon 2020 program of the European Commission.
\end{acknowledgments}

\section*{AUTHOR CONTRIBUTION}
E.S.M. conceived the idea of deep photonic networks. A.D.V. created the design framework with simulation, optimization, and layout capabilities. A.N.A. and K.G. developed and revised separate modules of the design framework. A.N.A. designed and simulated the individual devices. K.G. finalized the mask layout for fabrication. A.N.A performed the experimental characterization of the devices; and K.G. assisted the setup and experiments. E.S.M. supervised and coordinated the research. A.N.A and E.S.M. wrote the manuscript with contributions from all co-authors.

\section*{DATA AVAILABILITY}
The data that support the findings within this manuscript are available from the corresponding author upon reasonable request.


\section*{COMPETING INTEREST}
The authors declare no competing interests.

\def\bibsection{\section*{References}}
\bibliography{main}

\end{document}


\raggedbottom
\preprint{APS/123-QED}

\title{Deep Photonic Networks with Arbitrary and Broadband Functionality: \\ Supplemental Information}

\author{Ali Najjar Amiri}
\email{aamiri20@ku.edu.tr}
\author{Aycan Deniz Vit}
\author{Kazim Gorgulu}
\author{Emir Salih Magden}
\email{Corresponding author: esmagden@ku.edu.tr}
\affiliation{%
 Department of Electrical and Electronics Engineering, Koç University\\
Sariyer, Istanbul, 34450, Turkey
}%

\maketitle


\section{Modeling of Directional Couplers in MZI Interferometers}
For physical accuracy of the transfer matrix model employed, the complete optical transmission including any losses through the S-bends and directional couplers are incorporated in $T(\lambda)$, by using the optical response of the directional coupler obtained from 3D-FDTD simulations, as shown in Fig. S1.
\begin{figure*}[hb]
\includegraphics{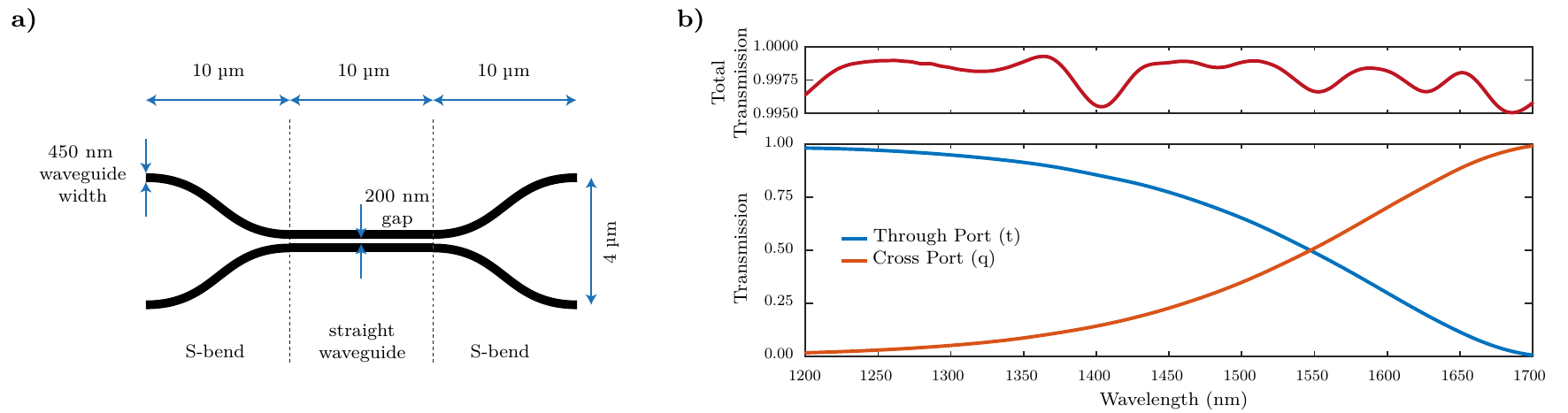}
\caption{\textbf{Diagram  and optical response of the directional coupler.} \textbf{(a)} Diagram of the directional coupler indicating important geometrical parameters including 10 \si{\um}-long straight coupling section with 200 \si{\nm} gap, 10 \si{\um}-long S-bends, 4 \si{\um} center-to-center waveguide separation between the top and the bottom arm at the start and end of the directional coupler, and 450 \si{\nm} waveguide width through the entire directional coupler. \textbf{(b)} Optical responses at the cross- and through-port of the directional coupler used in our deep photonic networks. Results were obtained from 3D-FDTD simulations performed with a maximum spatial discretization of 17 \si{\nm} in all three dimensions. Amplitude and phase information between 1.2 \si{\um} and 1.7 \si{\um} were implemented as automatic differentiation-compatible interpolations.}
\end{figure*}

Unlike the custom tapers in the interferometers, the directional couplers used in our networks are constructed to be all identical, with no trainable parameters. Due to the modular implementation of our design framework, it is possible to replace this specific directional coupler shown above with a different component, such as another directional coupler with a shorter or longer coupling length, one that uses bends with different geometries, or an entirely different coupler with a task-specific transfer function.

\section{INITIALIZATION AND OPTIMIZATION OF CUSTOM WAVEGUIDE TAPERS}
The photonic network aims to optimize the geometry of custom waveguide tapers and the resulting $\theta(\lambda)$ according to the objective function specified. However, even though a representative neural network model can optimize for specific $\theta(\lambda)$, this mathematical procedure can easily yield final widths and lengths that are not appropriate for an actual photonic device or our specific network architecture. To this end, in order to design our photonic devices, we build our physics-informed network models with specific initialization considerations, numerous physical boundaries, and regularization schemes imposed on the trainable parameters, as detailed below.

\subsection{Initialization and Constraints for Trainable Widths and Lengths}
As our demonstrations are on the 220 \si{\nm} SOI platform, the default waveguide width ($w_\mathrm{default}$) for the entire network including input/output waveguides, directional couplers, and bends, is chosen to be 450 \si{\nm}. As such, the final resulting widths in the custom tapers should also be relatively close to this default width, in order to ensure inherent fabrication compatibility of the designed devices, and also to minimize the excitation of higher-order modes as light propagates through these tapers.

Therefore, it is important that the array of custom waveguide widths are initialized relatively close to $w_\mathrm{default}$, which we implement by $w_\mathrm{initial} = w_\mathrm{default} + p_w \cdot \Delta w + w_\mathrm{offset}$ at the beginning of optimization. Here, $p_w$ is a  uniform random variable between -1 and 1, $\Delta w$ is a user-chosen maximum deviation amplitude, and $w_\mathrm{offset}$ is a constant initial offset. The designer can choose to specify an initial offset, which is especially useful if the parameter updates exhibit monotonic tendencies during optimization, which can be a common occurrence in some types of artificial neural networks \cite{pmlr-v28-sutskever13, alom2019state,hanin2018start}. Throughout optimization and calculation of the corresponding transfer functions, we also clip the trainable widths between $w_\mathrm{min} \le w \le w_\mathrm{max}$, in order to ensure sufficient mode confinement within the waveguide core, and to limit the development of strongly guided higher-order modes in the waveguides. For the networks we demonstrated, these parameters were typically chosen to be 5 \si{\nm} $\le \Delta w \le$ 20 \si{\nm}, 30 \si{\nm} $\le w_\mathrm{offset} \le$ 50 \si{\nm}, $w_\mathrm{min} =$ 400 \si{\nm}, and $w_\mathrm{max} =$ 520 \si{\nm}.\par

A similar implementation is used for the initialization of custom taper lengths with several minor differences. First, we initialize the lengths as $L_\mathrm{initial} = L_\mathrm{max} - p_L \cdot \Delta L$, where $p_L$ is a random variable between 0 and 1, and $\Delta L$ is a user-chosen maximum deviation amplitude. We similarly clip the trainable lengths between $L_\mathrm{min} \le L \le L_\mathrm{max}$ throughout the optimization. Additionally, if the resulting taper length is shorter than $L_\mathrm{max}$, in each iteration of the optimization procedure, we add a straight waveguide at the end of the taper with a length equal to $L_\mathrm{add} = L_\mathrm{max} - L$, ensuring that successive layers of MZIs align in the propagation direction. For demonstrated networks, these parameters were chosen to be $L_\mathrm{max} =$ 10 \si{\um} and $L_\mathrm{min} =$ 6 \si{\um}. With these parameter selections and 30 \si{\um}-long directional couplers, the resulting length of each MZI in our networks was 80 \si{\um}.

In addition to the initialization and imposed boundaries, the selection of the number of trainable widths per waveguide taper presents an important design consideration. By design, the ends of the custom tapers are fixed widths of $w_\mathrm{default}$; and the trainable widths are placed with equal spacing in between. As such, the number of trainable widths (together with $L_\mathrm{min}$ and $L_\mathrm{max}$ parameters) directly influences the resulting taper angle, which can result in potential propagation losses and/or excitation of any higher-order modes if the taper has insufficient length \cite{milton1977mode, fu2014efficient, zou2014short}. In our designs, we have chosen to place 5 uniformly-spaced trainable widths inside each waveguide taper, resulting in at least 1 \si{\um} spacing between them. Based on 3D-FDTD results, we find that a minimum taper length of $L_\mathrm{min} =$ 1 \si{\um} $\times\ (\xi + 1)$ is sufficiently long for the low-loss operation of tapers between $w_\mathrm{min} =$ 400 \si{\nm} and $w_\mathrm{max} =$ 520 \si{\nm}, where $\xi$ is the number of trainable widths in a single taper. The designer can also alternatively choose to place a greater number of trainable widths within each taper, and use a longer $L_\mathrm{min}$ parameter.
\subsection{Regularizers}
In addition to the initialization and constraints detailed above, two major regularization schemes are implemented in order to control the trainable widths through the iterative optimization. The first regularizer aims to reduce the difference between consecutive widths in a single waveguide taper, in order to prevent abrupt changes in width and potential resulting propagation losses. The second regularizer aims to limit the difference between any one of trainable widths and a reference width $w_\mathrm{ref}$ (which may be chosen to be the same as $w_\mathrm{default}$), in order to maintain the optimized variables within the relative vicinity of this reference width. Both regularizers are calculated as L2-norms of the corresponding error vectors as 
\begin{equation}
    P_{1} = \sum_\mathrm{all \ tapers}\sum_{i=1}^{\xi- 1} (w_{i} - w_{i+1})^2 
\end{equation}
and
\begin{equation}
    P_{2} = \sum_\mathrm{all \ tapers}\sum_{i=1}^{\xi} (w_{i} - w_\mathrm{ref})^2 
\end{equation}
where $w_i$ are the individual trainable widths in each custom taper in a given photonic network. An overall regularization is computed as $P = \alpha_1 P_1+\alpha_2 P_2$, where coefficients $\alpha_1$ and $\alpha_2$ are used to independently control the contribution strength of the two regularizers. The accumulated contribution of the regularizers is then implemented as an artificially-introduced loss at the end of the network, after which the overall objective function is calculated as
\begin{equation}
    J(\mathbit{x})=\frac{1}{Q}\sum_{\lambda}\mid T_\mathrm{calculated}(\lambda , \mathbit{x}) e^{-P} -T_\mathrm{target}(\lambda)\mid ^2
\end{equation}
where $Q$ is the number of wavelengths and $\mathbit{x}$ are design parameters including widths and lengths of the custom tapers. For the networks demonstrated here, we used $\alpha_1 = 3\times{10}^{-4}$ and $\alpha_2 = 1\times{10}^{-4}$. Depending on the photonic capabilities required, the designers may freely experiment with various contribution strengths and examine resulting taper profiles. As expected, stronger regularizations result in waveguides with more slowly varying widths, but also reduce the capability of the photonic network to replicate arbitrary transfer functions, due to the reduced effective degrees of freedom. Thanks to the computational efficiency of our network implementation, experimenting with different regularization contributions and comparing resulting taper structures is trivial from a designer’s perspective.

\section{HYPERPARAMETER CONSIDERATIONS FOR DEEP PHOTONIC NETWORKS}
Similar to many other machine learning models, the selection of appropriate hyperparameters has key implications on the capability and scalability of our deep photonic networks, as well as the computational efficiency of their physics-informed, representative, artificial neural network models. In this section, we provide details on the effects of the number of interferometric layers and the number of trainable parameters per MZI on the final performance of our photonic networks.
\subsection{Number of Interferometric Layers}
The number of interferometric layers (the network depth) directly influences the number of trainable parameters, and is one of the key hyperparameters of our photonic networks. As demonstrated in Fig. 5, this network depth strongly affects the final performance and the tolerance against fabrication imperfections of the resulting photonic device. In Fig. S2, we perform similar investigations for the 75/25 power splitter and the spectral duplexer devices by comparing the final performance and robustness of 10 different devices designed from $M = 2$ to $M = 60$ layers, with the resulting number of trainable parameters ranging between 48 and 1440.

In Fig. S2(a) and (c), we plot the calculated errors for each type of photonic device and observe an initial decrease in this error with the increasing number of layers due to increased network capability. Minimum errors are reached around $M = 3$ to 5 layers for the 75/25 splitter, and with $M = 9$ to 14 layers for the duplexer. As the duplexer performs a more complicated spectral functionality, it requires a deeper, and therefore more capable network, with greater degrees of freedom. However, deeper networks also suffer more strongly from fabrication imperfections resulting from phase errors, and other propagation losses including those in directional couplers and S-bends. This can be verified from the general profile of the errors as a function of over-etch/under-etch offset $\Delta w$ plotted in Fig. S2(b) and (d), where shorter networks generally demonstrate better (flatter) performance under strong etch offsets reaching $±$20 \si{\nm}. Specifically for the 75/25 splitter, while the minimum error is reached with a 4-layer network, this 4-layer network exhibits significantly worse fabrication tolerance under $±$20 \si{\nm} etch offsets, in comparison with the 3-layer network. Likewise, for the spectral duplexer, while the minimum error is reached with a 10-layer network, the fabrication tolerance of this 10-layer network is significantly inferior in comparison to the 6-layer and 7-layer networks. In comparing our devices, we note that the duplexer demonstrates better fabrication tolerance in these simulations than both of the power splitters. This difference can be attributed to the final error of the duplexer ($2 \times {10}^{-4}$) already being 1-2 orders of magnitude greater than the final errors for the power splitters (between $7 \times {10}^{-6}$ and $2 \times {10}^{-5}$).

In general and also for our specific designs here, we consider both the final error and the fabrication tolerance of the resulting photonic networks. In cases where devices with different numbers of layers exhibit closely similar performances in calculated errors, we choose the device with a smaller number of layers in order to ensure a greater degree of robustness against fabrication variations, and also for compact device footprints.
\begin{figure*}
\includegraphics{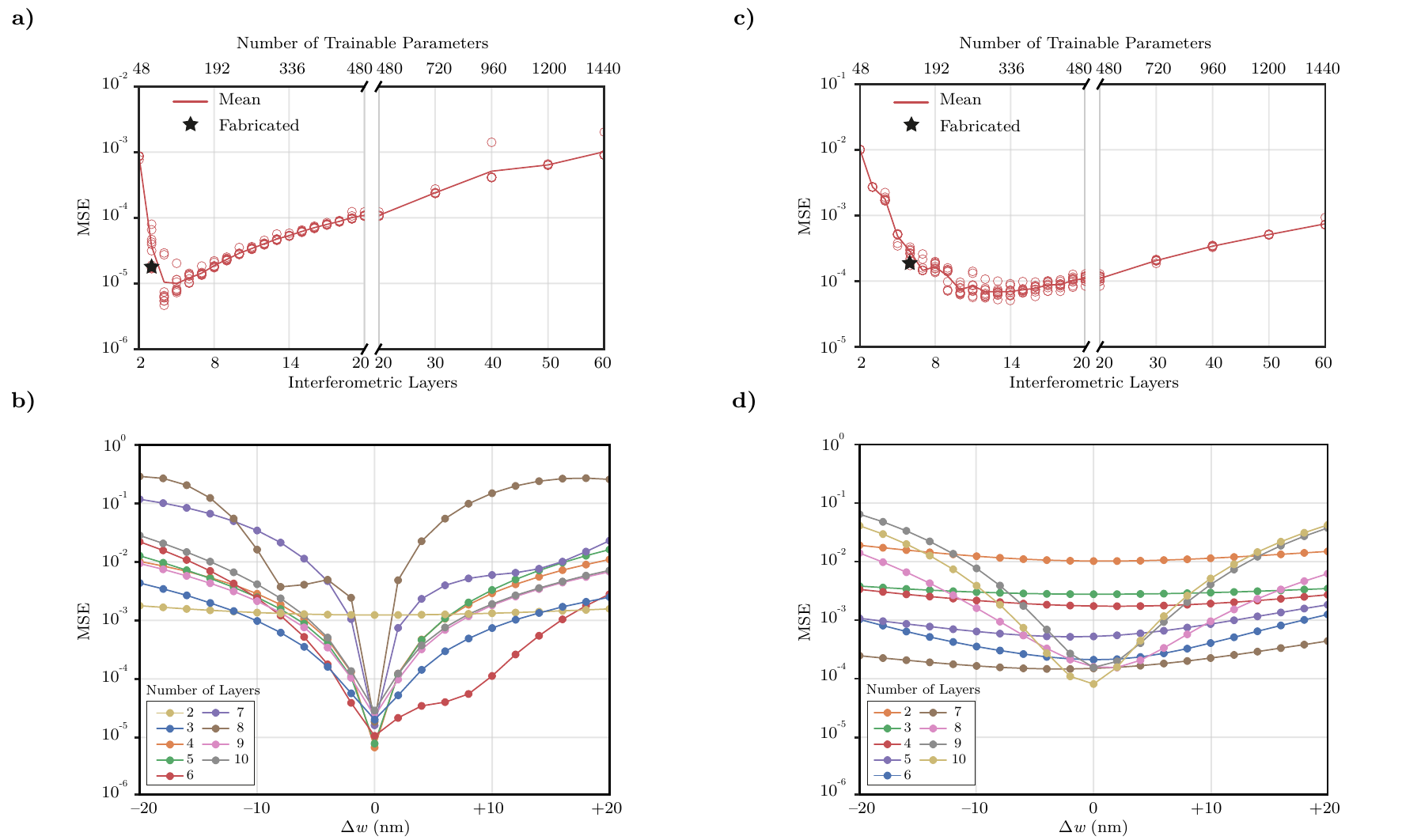}
\caption{\textbf{Influence of network size on final device performance.} Mean squared error for \textbf{(a)} 75/25 power splitter, and \textbf{(c)} spectral duplexer deep photonic networks as functions of the number of interferometric layers. The plotted error includes propagation loss in the directional couplers and the S-bends extracted from their 3D-FDTD simulations. For each network size, ten different randomly-initialized devices are optimized and depicted with red circles. Robustness against fabrication variations for \textbf{(b)} 75/25 power splitter, and \textbf{(d)} spectral duplexer. A greater number of interferometric layers initially reduces the calculated errors. However, longer devices suffer from stronger deviations from target functionality under fabrication variations.}
\end{figure*}

\subsection{Number of Trainable Parameters}
The trainable parameters consist of the widths and the length for each custom taper in our photonic networks, as illustrated in Fig. 2. Specifically, each MZI in the network is made from four waveguide tapers built from trainable widths, four trainable lengths (one for each taper), and two fixed (non-trainable) directional couplers. By design, the network depth directly determines the number of trainable lengths. However, the number of trainable widths per custom taper is a designer-specified parameter. In the presented devices, we used five trainable widths for each taper, yielding a total of $4 \times (5 + 1) = 24$ trainable parameters per MZI. For each taper, these constituent widths and the taper length are used to calculate the accumulated phase, as shown in Fig. 1. As discussed in Supplementary Section 2A, we generally limit the number of trainable widths as $\xi \le (L_\mathrm{min} - 1\ \si{\um}$)/1\ \si{\um}, in order to ensure slowly-varying waveguide geometries in the propagation direction for minimizing unwanted loss. For the 10 \si{\um} maximum taper lengths in our devices, this results in the recommended number of trainable widths being less than or equal to 9. However, within this bound, the number of trainable widths still remains a user-specifiable design choice. For our devices demonstrated here, we used $\xi = 5$ trainable widths per taper, as we find that there is a negligible added benefit beyond five trainable widths, since the required optimized phase profiles can already be reached in the custom tapers. Additionally, the number of trainable widths $\xi > 5$ may also result in slightly longer optimization times, with no improvements in device performance. 

\section{computational performance and time requirements for network optimization}
One of the most important advantages of our deep photonic network design framework is its ability to replace resource-extensive and time-consuming 3D-FDTD simulations with physically accurate, yet computationally efficient scattering matrix calculations. These computational requirements for underlying simulations are especially critical in design tasks where device parameters are repeatedly modified, and the device response is simulated in an iterative manner. Our design framework achieves computationally efficient and highly scalable operation through its use of state-of-the-art machine learning software and hardware infrastructure. 

In Fig. S3, we demonstrate this computational performance for the design of several power splitters with various numbers and configurations of output ports, using a single Tesla V100 GPU. Specifically, we consider 2-port, 4-port, and 8-port broadband power splitters in evenly distributed and randomly distributed output power configurations, optimized at 32 evenly-spaced wavelengths between 1400-1600 \si{\nm}. In Fig. S3(a), we plot the total time required for the convergence of each device as a function of the network depth. A relative convergence criterion of  
\begin{equation}
    \frac{\mid J(\mathbit{x})_\mathrm{current} - J(\mathbit{x})_\mathrm{previous}\mid}{\mathrm{max}(J(\mathbit{x})_\mathrm{current}, J(\mathbit{x})_\mathrm{previous})} < 10^{-3}
\end{equation}
was used for all devices where $J(\mathbit{x})$ was defined in Eq.(3) above. As expected, the total optimization time scales with the number of interferometric layers ($M$) and the number of network outputs ($N$). As a funcation of these two parameters and the number of trainable widths per custom waveguide taper, the total number of trainable network parameters is given by
\begin{equation}
  \begin{cases}
    4(\xi + 1)M, & \text{$N = 2$}\\
    4(\xi + 1)(\lceil \frac{M}{2} \rceil \lfloor \frac{N}{2} \rfloor + \lfloor \frac{M}{2} \rfloor \lfloor \frac{N - 1}{2} \rfloor), & \text{$N \geq 3.$}
  \end{cases}
\end{equation}
However, even for photonic networks as deep as 10 layers (800 \si{\um} device length), with 8 separate output ports, and a highly complicated photonic functionality like random and broadband output power splitting, the entire optimization procedure is completed in less than 12 minutes. This result presents superior scalability and multiple orders of magnitude advancements over optimization procedures based on 3D-FDTD simulations \cite{piggott2015inverse,lu2013nanophotonic,jia2018inverse,zhang2021scalable,zhang2022experimental}. For smaller photonic networks with simpler functionalities, our optimizations complete in less than approximately 1 minute for networks of up to four layers. This capability of designing devices with physically accurate simulations within several minutes gives photonic designers the crucial ability to easily iterate through different versions of their photonic networks, and investigate device performance using different hyperparameters in their designs. The scalability of our design framework is also demonstrated by the per-iteration-time for each network optimization plotted in Fig. S3(b). Here, an iteration is defined as the computational operation of updating the network parameters once, after calculating $J(\mathbit{x})$ and the required gradient $\nabla_x J$ using all 32 wavelengths in a parallelized manner. As a function of the network depth (and therefore the number of trainable parameters), the time required for each iteration on average ranges between a few milliseconds and a few tens of milliseconds. In addition to the modularity of our photonic network construction, this performance was also partly enabled by the use open-source deep learning software libraries JAX \cite{jax2018github} and Trax \cite{Trax2020}. These libraries allow automatic differentiation-compatible computations to be performed by CPU, GPU, or TPU accelerator hardware in a parallelized manner through just-in-time (JIT) compilation \cite{aycock2003brief} and execution of required function calls. For our photonic networks, after a model is created, a certain preparation time is necessary in order for JIT compilation to complete, before iterative optimization can begin. For specific optimizations shown in Fig S3, this preparation time ranged between 8 and 650 seconds for 1-layer to 10-layer deep networks. This is included in the total optimization durations plotted in Fig. S3(a), but excluded from the per-iteration time plotted in Fig. S3(b). 

\begin{figure*}[hb]
\includegraphics{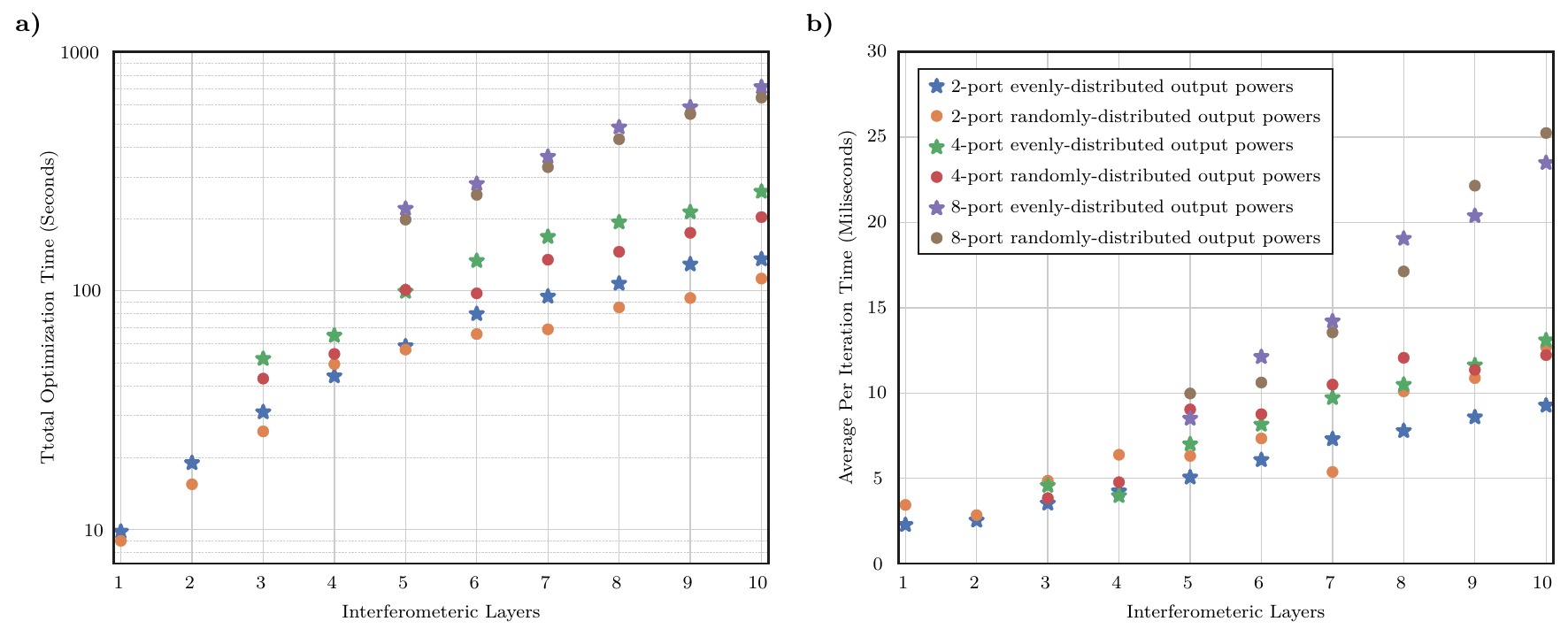}
\caption{\textbf{Computational performance of the design framework for deep photonic networks as arbitrary power splitters.} \textbf{(a)} Total optimization time for networks with different number of interferometric layers. \textbf{(b)} Average time per optimization iteration. Networks up to 10-layers deep with the number of output ports $N =$ 2, 4, and 8 are optimized for broadband power splitting using 32 wavelengths between 1400 \si{\nm} and 1600 \si{\nm}. Separate photonic networks are designed with evenly distributed output powers, and randomly distributed output powers. Light is injected at the input number $\lfloor \frac{N}{2} \rfloor$ for all devices. All optimizations were performed using an open-source, end-to-end deep learning software library \cite{jax2018github, Trax2020}, adaptive moment estimation optimizer \cite{kingma2014adam}, and a single Tesla V100 GPU. Optimizations were completed in several hundred iterations for each device, using a convergence criterion of $10^{-3}$ relative change in the overall objective $J(\mathbit{x})$.}
\end{figure*}

\def\bibsection{\section*{References}}
\bibliography{supplement}